\documentstyle[12pt]{article}
\begin{document}

\title{{\bf Multiple scattering}}
\author{
Ricardo Garc\'\i a-Pelayo\thanks{%
E-mail: rgpelayo@fisfun.uned.es} \\
\\
Departamento de F\'{\i}sica Fundamental, UNED \\
Apdo. de Correos 60141 \\
Madrid 28080, Spain}
\date{}
\maketitle

\centerline{\bf{Abstract}}

\medskip

{\small{The purpose of this work is to find the time dependent distributions of directions and positions of a particle that undergoes multiple elastic scattering. The angular cross section is given and the scatterers are randomly placed. 

The distribution of directions is found. As for the second distribution we find an exact expression for isotropic cross section in 2 dimensions. For the same cross section we find its Fourier-Laplace transform in 3 dimensions. For the general case we devise a method to compute the Fourier-Laplace transform with arbitrary precision.

This work is based not on the Boltzmann transport equation but on an integral equation formulation of the problem. The results are general in the sense that any initial condition is a linear combination of the cases considered in this article.}}

\bigskip

Key words: Boltzmann transport equation, multiple scattering, integral equations, Lorentz gases.

\bigskip
PACS-96 numbers: 02.30.Rz, 05.20.Dd, 05.40.+j, 05.60.+w.

\section{Applications and overview}

As a quantitative analytical technique multiple scattering is used to study clouds [1], pollution, chemichals in solution [2], radiation in the human body [3,4,5], etc... A great effort has been made to study the scattering of laser beams by clouds. To this end some authors [1] have developped numerical stochastic models, based on the same general philosophy as ours, which have been shown to be equivalent to the transport equation which is more commonly used. Simulations are a major tool in the study of multiple scattering [1,6], but real situations are computationally so demanding that numerical methods based on analitical understanding of the problem can be necessary. 

In Physics multiple scattering is obviously related to Lorentz gases [7] and to diffusion of neutrons in nuclear reactors [8,9]. It has many applications in solid state [6]. In most situations the particle being scattered is a photon or an electron. Here we shall work with a scalar particle, but the methods can be generalized for the case of internal degrees of freedom as long as interference effects remain negligible.

In the 2nd section we find the directions taken by the particles of a beam which encounters randomly distributed diffusing centers. Conversely the differential cross section of the particle-scatterer interaction can be solved in terms of the distribution of directions. Asymptotics are also discussed. The results of the 2nd section are experimentally applicable as long as the detectors are placed at a distance to the medium to be studied which is much larger than the diameter of the medium. In the 3rd section the motion of the center of mass is found. In the 4th section we study the case of isotropic cross section. The probability density for the particles is found explicitely in 2 dimensions and in 3 dimensions its Fourier-Laplace transform is found. In the 5th section we show how to calculate the Fourier-Laplace transform of the probability density with arbitrary precision in the general anisotropic case for both 2 and 3 dimensions.

The initial condition for all the results in this article is a beam of particles. This is the general case in the sense that any initial condition can be decomposed in terms of beams. The mathematical setting of this work is that of integral equations. For some instances this work provides an alternative to the Boltzmann equation for the study of transport processes.

\section{Diffusion of directions}

Consider the repeated elastic scattering of a point-like particle by randomly distributed recoilless diffusing centers. We assume that the scattering itself takes place in a region of negligible volume (or area) in the sense that the trajectory of the particle can be well approximated by a sequence of straight line segments. Let $f$ be the angular cross section, where $f$ is a function of the angle. If $f$ is not equal to a constant, i. e., if the scattering is not isotropic, then a statistical description of the multiple scattering becomes difficult because the process is not Markovian in the physical space. However, the stochastic process for the {\it{directions}} is Markovian. In fact if, in $d$ dimensions, we represent the direction along which the particle is moving by a point on $S^{d-1}$, then the probability of transition from one point of $S^{d-1}$ to another will depend only on the angle between them, so that convolution techniques can be applied. We will show this in 2 and then in 3 dimensions.

In 2 dimensions let $F(t, \varphi)$ be the probability density for the particle to be moving along the direction $\varphi$ at time $t$. ${\tilde{}}$ will denote Fourier transform and $\overline{\varphi}$ will be the Fourier frequency associated to the angle $\varphi$. If $\lambda$ is the probability of collision per unit time then in an infinitesimal time $dt$ the probability density $F$ will change to:

\begin{equation}
F(t+dt, \varphi) = (1 - \lambda\ dt) F(t, \varphi) + \lambda\ dt \int d \varphi'\ F(t, \varphi') f(\varphi - \varphi'), 
\end{equation}

\noindent
The Fourier transform of the above equation is: 

\begin{equation}
{\tilde F}(t+dt, \overline{\varphi}) = (1 - \lambda\ dt) {\tilde F}(t, \overline{\varphi}) + \lambda\ dt\ {\tilde F}(t, \overline{\varphi}) {\tilde f } (\overline{\varphi}) = {\tilde F}(t, \overline{\varphi}) +  \lambda\ dt\ {\tilde F}(t, \overline{\varphi}) ({\tilde f}(\overline{\varphi}) - 1).
\end{equation}

\noindent
This yields the differential equation:

\begin{equation}
{\partial \over \partial t} {\rm {ln}} {\tilde F} (t, \overline{\varphi}) = \lambda [{\tilde f } (\overline{\varphi})-1], 
\end{equation}

\noindent
whose solution is:

\begin{equation}
{\tilde F} (t, \overline{\varphi}) = {\tilde F} (0, \overline{\varphi})\ \exp \lambda [{\tilde f } (\overline{\varphi})-1] t. 
\end{equation}

An analogous reasoning shows that when $\lambda$ or $f$ are time dependent:

\begin{equation}
{\tilde F} (t, \overline{\varphi}) = {\tilde F} (0, \overline{\varphi})\ \exp \int_{0}^{t} dt \ \lambda(t) \ [{\tilde f } (t, \overline{\varphi})-1].  
\end{equation}

These formulae are valid for any cross section $f$. In particular left-right symmetry ($f(\varphi) = f(2 \pi- \varphi)$) is not required.

When all particles move along the direction $\varphi=0$ at $t=0$, ${\tilde F} (0, \overline{\varphi})=1$ and the above formulae simplify. This kind of initial condition, which we shall call stream initial condition, is important because an arbitrary initial condition can be constructed in terms of it. In 3 dimensions the stream initial condition will be a beam of particles moving towards the positive direction of the $z$ axis, due to the definition of polar coordinates in 3 dimensions.

In 3 dimensions the equation corresponding to (1) is:

\begin{equation}
F(t+dt, \Omega) = (1 - \lambda\ dt)\ F(t, \Omega) + \lambda\ dt \int d \Omega'\ F(t, \Omega')\ f(\Omega - \Omega'), 
\end{equation}

\noindent
where $\Omega - \Omega'$ is the angle between the directions $\Omega$ and $\Omega'$. Here we are going to assume that the scattering probability $f$ depends only on the polar angle $\theta$, or, in other words, that if the particle undergoing scattering has internal (pseudo)vectorial degrees of freedom (such as the polarization of a photon), these are not involved in the scattering or are averaged over. We also assume, for simplicity's sake, that the initial condition $F(0, \Omega)$ is independent of $\varphi$. Then $F(t, \Omega)$ is also going to be $\varphi$ independent and both $F(t, \Omega)$ and $f(\Omega)$ can be expanded in terms of Legendre polynomials $P_{l}$ as follows (see [10] or [11]):

\begin{equation}
F(t, \Omega) = \sum_{l=0}^{\infty} \sqrt{{2l+1 \over 4 \pi}}\ F_{l}(t)\ Y_{l}^{0} (\theta)
\end{equation}

\noindent
and

\begin{equation}
f(\Omega) = \sum_{l=0}^{\infty} \sqrt{{2l+1 \over 4 \pi}}\  f_{l}\ Y_{l}^{0} (\theta),
\end{equation}

\noindent
where 

\begin{equation}
Y_{l}^{0} (\theta) = \sqrt{{2l+1 \over 4 \pi}}\ P_{l} (\cos \theta)
\end{equation}

\noindent
are the spherical harmonics and

\begin{equation}
F_{l}(t) = \int d\Omega\ \sqrt{{4 \pi \over 2l+1}}\ F(t, \Omega) Y_{l}^{0} (\Omega),
\end{equation}

\noindent
and a similar formula for $f_{l}$.

If we now substitute these expansions in the convolution product in (6) and use the sum formula for Legendre polynomials,
 
\begin{equation}
Y_{l}^{0} (\Omega-\Omega') = \sqrt{{4 \pi \over 2l+1}} \sum_{m=-l}^{l} Y_{l}^{m*}(\Omega) Y_{l}^{m}(\Omega'),
\end{equation}

\noindent
and 

\begin{equation}
\int d\Omega\ Y_{l}^{m*} (\Omega) Y_{l'}^{m'} (\Omega) = \delta_{ll'} \delta_{mm'}, 
\end{equation}

\noindent
we obtain:

\begin{equation}
\int d \Omega'\ F(t, \Omega')\ f(\Omega - \Omega') = \sum_{l=0}^{\infty} \sqrt{{2l+1 \over 4 \pi}}\ F_{l}(t)\ f_{l}\ Y_{l}^{0} (\Omega).
\end{equation}

\noindent
Proceeding as in the 2-dimensional case we obtain the time evolution of the coefficients $F_{l}(t)$:

\begin{equation}
F_{l}(t) = F_{l}(0)\ \exp \lambda (f_{l}-1) t,\ \ \ \ l=0, 1, 2,...,
\end{equation}

\noindent
as well as an equation analogous to (5). For the $z$-stream initial condition, $F_{l}(0) = 1$ for $l=0, 1, 2,...$. Different derivations of eq. (14) have been given by Goudsmit and Saunderson ([12],[13]) and by Lewis [14] (these derivations are also reviewed in [6]).

Note that the fact that $f(\theta)$ is arbitrary allows us to treat cross sections like the one of the Rayleigh scattering, which a Fokker-Planck equation on the sphere (e. g. the Debye rotational diffusion equation [15]) can not approximate.

It is intuitive to think that asymptotically the directions will be uniformly distributed. From

\begin{equation}
F(t, \varphi)= {1 \over 2 \pi} \sum_{\overline{\varphi} \epsilon Z} {\tilde F}(0, \overline{\varphi})\ e^{\lambda [{\tilde f}(\overline{\varphi})-1] t }\ e^{i \overline{\varphi} \varphi},
\end{equation}

\noindent 
and

\begin{equation}
\tilde f (\overline{\varphi}) = \int_{0}^{2 \pi} d\varphi \ f(\varphi) \ e^{-i \overline{\varphi} \varphi},
\end{equation}

\noindent
we see that unless $f(\varphi)$ is a set of Dirac deltas placed at the vertices of a regular polygon, our final distribution in 2 dimensions is indeed going to be uniform. 

In 3 dimensions normalization implies $f_{0}=1$. We now show that the other coefficients, $f_{l},\ \ l=1, 2,...$, are smaller than 1:

\begin{equation}
f_{l} = \int d\Omega\ \sqrt{{4 \pi \over 2l+1}} f(\Omega) Y_{l}^{0}(\Omega) =
2 \pi \int_{-1}^{1} d(\cos \theta)\ f(\theta)\ P_{l}(\cos \theta).
\end{equation}

\noindent
The Legendre polynomials take absolute values always smaller than 1 except for $\cos \theta = \pm 1$, where the even ones take the value 1 and the odd ones take the value $\pm 1$. Thus if $f(\theta) \neq \alpha\ \delta(\theta) + \overline{\alpha}\ \delta(\theta - \pi)$, where $\alpha + \overline{\alpha} = 1$, then

\begin{equation}
f_{l} < 2 \pi \int_{-1}^{1} d(\cos \theta)\ f(\theta) = \int d\Omega\ f(\Omega) = 1,\ \ \ \ \ l=1, 2,...,
\end{equation}

\noindent
and the time evolution (14) implies that in 3 dimensions the final distribution $F(\infty, \Omega)$ will be the uniform distribution.

To measure quantitatively how the distribution approaches its final state, we choose the $L^{2}$ distance between the current distribution and the final one. In 2 dimensions this is:

\begin{equation}
\int_{0}^{2 \pi} d\varphi \ \bigg{(}F(t, \varphi)-{1 \over 2 \pi} \bigg{)}^{2}.
\end{equation}

The Fourier transform is an isometry of the $L^{2}$ norm, so that the above distance is equal to:

\begin{equation}
{1 \over 2 \pi} \sum_{\overline{\varphi} \neq 0} e^{2 \lambda [{\tilde f(\overline{\varphi}) }-1] t }.
\end{equation}

The time dependence of the second cumulant of the probability density for finding a particle which is suffering multiple scattering tends to be linear as $t \rightarrow \infty$. By studying the approach to this behaviour Bogu\~{n}\'a {\it{et alii}} [16] found the above formula with the sum restricted to $\overline{\varphi} = \pm 1$. Higher terms in (20) are related to higher moments.

In 3 dimensions the $L^{2}$ distance between the current distribution, $F(t, \Omega)$, and the final one, ${1 \over 4 \pi}$, can be expressed in a form analogous to (20) using expansion (7), the orthonormality relation (12) and the evolution equation (14). The final expression for the $L^{2}$ distance is:

\begin{equation}
\sum_{l=1}^{\infty} {2l+1 \over 4 \pi}\ F_{l}(0)^2\ e^{2 \lambda (f_{l}-1) t}
\end{equation}
\bigskip

\section{Motion of the center of mass}

If the cross section $f$ is not ballistic then at every collision some of the initial momentum of the particle is transfered to the scatterer. As a result the velocity of the center of mass goes exponentially to zero. It is interesting to find how far the center of mass will go before (at $t=\infty)$ it stops.  To study this note that in 2 dimensions for a particle (which we assume of unit mass) impinging along the direction $\varphi$ upon a scatterer, its momentum changes from $(v \cos \varphi, v \sin \varphi)$ to, on the average, $(v \int_{0}^{2 \pi} d\varphi' f(\varphi'-\varphi) \cos \varphi', v \int_{0}^{2 \pi} d\varphi' f(\varphi'-\varphi) \sin \varphi')$. Thus Newton's law for the center of mass reads:

\begin{equation}
{{d\vec p} \over dt} = \lambda v
$$

$$ 
\int_{0}^{2 \pi} d\varphi\ F(t, \varphi)
\Bigg{(}
\int_{0}^{2 \pi} d\varphi' f(\varphi'-\varphi) \cos \varphi'- \cos \varphi,
\int_{0}^{2 \pi} d\varphi' f(\varphi'-\varphi) \sin \varphi'- \sin \varphi
\Bigg{)},
\end{equation}

\noindent
or, using (15),

\begin{equation}
{{d\vec p} \over dt} = {\lambda v \over 2 \pi} \sum_{\overline{\varphi} \epsilon Z} {\tilde F}(0, \overline{\varphi})\ e^{\lambda [{\tilde f}(\overline{\varphi})-1] t}
$$

$$ 
\int_{0}^{2 \pi} d\varphi\  e^{i \mu \varphi}
\Bigg{(}
\int_{0}^{2 \pi} d\varphi' f(\varphi'-\varphi) \cos \varphi'- \cos \varphi,
\int_{0}^{2 \pi} d\varphi' f(\varphi'-\varphi) \sin \varphi'- \sin \varphi
\Bigg{)} = 
$$

$$
{\lambda v \over 2 \pi} \sum_{\overline{\varphi} \epsilon Z} {\tilde F}(0, \overline{\varphi})\ e^{\lambda [{\tilde f}(\overline{\varphi})-1] t}
$$

$$ 
\Bigg{(} {\tilde f}(\overline{\varphi}) \pi [\delta_{\overline{\varphi}, -1} + \delta_{\overline{\varphi}, 1}] -  \pi [\delta_{\overline{\varphi}, -1} + \delta_{\overline{\varphi}, 1}], 
{\tilde f}(\overline{\varphi}) {\pi \over i} [\delta_{\overline{\varphi}, -1} - \delta_{\overline{\varphi}, 1}] -  {\pi \over i} [\delta_{\overline{\varphi}, -1} - \delta_{\overline{\varphi}, 1}]
\Bigg{)} = 
$$

$$
{v \over 2} \sum_{\overline{\varphi} \epsilon Z} {\tilde F}(0, \overline{\varphi})\ {d \over dt} e^{\lambda [{\tilde f}(\overline{\varphi})-1] t} 
\Bigg{(} [\delta_{\overline{\varphi}, -1} + \delta_{\overline{\varphi}, 1}], 
-i\ [\delta_{\overline{\varphi}, -1} - \delta_{\overline{\varphi}, 1}]
\Bigg{)}
\end{equation}

\noindent
Since the initial condition $F(0, t)$ and the scattering cross section $f(\varphi)$ are real valued functions, ${\tilde F}(0, -1) = {\tilde F}(0, 1)^*$ and ${\tilde f}(-1)={\tilde f}(1)^*$. Therefore

\begin{equation}
{{d\vec p} \over dt} = 
{\rm {Re}} \Big{[} 
v {\tilde F}(0, 1)\ {d \over dt} e^{\lambda [{\tilde f}(1)-1] t}(1, i) \Big{]}
\end{equation}

For the initial condition ${\vec p} (0) = (v \int_{0}^{2 \pi} d\varphi\ F(0, \varphi) \cos \varphi, v \int_{0}^{2 \pi} d\varphi\ F(0, \varphi) \sin \varphi)$ the solution to this differential equation is:

\begin{equation}
{\vec p} (t) = 
{\rm {Re}} \Big{[} 
v {\tilde F}(0, 1)\ e^{\lambda [{\tilde f}(1)-1] t}(1, i) \Big{]}
\end{equation}

\noindent
In particular for the stream initial condition and symmetric cross section,  

\begin{equation}
{\vec p} (t) = v\ e^{\lambda [{\tilde f}(1)-1] t}\ \vec i.
\end{equation}

Finally, for ${\vec x} (0) = \vec 0$ a last integration yields:

\begin{equation}
{\vec x} (t) = 
{\rm {Re}} \Big{[} 
v {\tilde F}(0, 1)\ {(e^{\lambda [{\tilde f}(1)-1] t} - 1) \over {\lambda [{\tilde f}(1)-1]}} (1, i) \Big{]}.
\end{equation}

\noindent
For the stream case with symmetric cross section this implies that the center of mass penetrates a distance  

\begin{equation}
|{\vec x} (\infty)| = {v \over {\lambda[1- {\tilde f}(1)]}}.
\end{equation}

\noindent
into the scattering medium.

In 3 dimensions for a particle of unit mass impinging along the direction $\Omega\equiv(\theta, \varphi)$ upon a scatterer, its momentum changes from $(v \sin \theta \cos \varphi, v \sin \theta \sin \varphi, v \cos \theta)$ to, on the average, $(v \int_{0}^{\pi} d\theta' \sin \theta' \int_{0}^{2 \pi} d\varphi' f(\theta'-\theta, \varphi'-\varphi) \sin \theta' \cos \varphi', v \int_{0}^{\pi} d\theta' \sin \theta' \int_{0}^{2 \pi} d\varphi' f(\theta'-\theta, \varphi'-\varphi) \sin \theta' \sin \varphi', v \int_{0}^{\pi} d\theta' \sin \theta' \int_{0}^{2 \pi} d\varphi' f(\theta'-\theta, \varphi'-\varphi) \cos \theta')$. Thus Newton's law for the center of mass reads: 

\begin{equation}
{{d\vec p} \over dt} = \lambda v \int_{0}^{\pi} d\theta \sin \theta \int_{0}^{2 \pi} d\varphi\ F(t, \theta, \varphi)
$$

$$ 
\Bigg{(}\int_{0}^{\pi} d\theta' \sin \theta' \int_{0}^{2 \pi} d\varphi' f(\theta'-\theta, \varphi'-\varphi) \sin \theta' \cos \varphi'-\sin \theta \cos \varphi,
$$

$$ 
\int_{0}^{\pi} d\theta' \sin \theta' \int_{0}^{2 \pi} d\varphi' f(\theta'-\theta, \varphi'-\varphi) \sin \theta' \sin \varphi'-\sin \theta \sin \varphi,
$$

$$ 
\int_{0}^{\pi} d\theta' \sin \theta' \int_{0}^{2 \pi} d\varphi' f(\theta'-\theta, \varphi'-\varphi) \cos \theta'- \cos \theta\Bigg{)}.
\end{equation}

As in section 2 we restrict ourselves to $F$ and $f$ functions of $\theta$ only. Then only the $z$ component of the above vector is non zero. Using the expression (14) for the time evolution of the Legendre polynomials coefficients of $F(t, \theta)$ we obtain:                            

\begin{equation}
{{d\vec p} \over dt} = \lambda v \int_{0}^{\pi} d\theta \sin \theta \int_{0}^{2 \pi} d\varphi\ \sum_{l=0}^{\infty} \sqrt{{2l+1 \over 4 \pi}}\ F_{l}(0)  e^{\lambda [f_{l}-1] t} Y_{l}^{0} (\theta)
$$

$$
\big{(}0, 0, \int_{0}^{\pi} d\theta' \sin \theta' \int_{0}^{2 \pi} d\varphi' f(\theta'-\theta) \cos \theta'- \cos \theta\big{)}
\end{equation}

Using the summation formula for Legendre polynomials (11) and the orthonormality of the spherical harmonics (12) we obtain

\begin{equation}
{\vec p} (t) = v F_{1}(0) e^{\lambda [f_{1}-1] t} \vec k.
\end{equation}

\noindent
For the $z$-stream initial condition, $F(0, \Omega)={\delta(\cos \theta-1) \over 2 \pi}$, $F_{1}(0)=1$, and ${\vec p} (t) = v e^{\lambda [f_{1}-1] t} \vec k$.

From (31), 

\begin{equation}
{\vec x} (t) = 
v F_{1}(0)\ {(e^{\lambda [f_{1}-1] t} - 1) \over {\lambda [f_{1}-1]}}\ \vec k.
\end{equation}

\noindent
and

\begin{equation}
|{\vec x} (\infty)| = {v F_{1}(0) \over {\lambda[1-f_{1}]}}\ \vec k.
\end{equation}

\noindent
For the $z$-stream  case we obtain a formula similar to the 2-dimensional one:

\begin{equation}
|{\vec x} (\infty)| = {v \over {\lambda[1-f_{1}]}}\ \vec k.
\end{equation}

\section{Isotropic scattering}

When the scattering is isotropic there is a direct way of obtaining the $x$-stream probability density ($\rho_{x}$) in terms of the probability density for isotropic initial conditions ($\rho$):

\begin{equation}
\rho_{x}(t, \vec x) = e^{-\lambda t} \delta(x-vt) \delta(y) + \lambda \int_{0}^{t} dt'\ 
e^{-\lambda t'} \rho(t-t', |\vec x - v t'\vec i|).
\end{equation}

The rationale for the above equation is that at each collision the stream gives rise to an isotropic probability cloud.

For the 2-dimensional case $\rho$ is known [17]: 

\begin{equation}
\rho(t, \vec x) = e^{-\lambda t} \Bigg{[} {\delta(r-vt) \over 2 \pi r} + 
{\lambda \over{2 \pi v  \sqrt {(vt)^{2}-r^2}}} \exp \Big{(} {{\lambda \over v} \sqrt {(vt)^{2}-r^2}} \Big{)}  H(vt-r)\Bigg{]},
\end{equation}

\noindent
where $H$ is the Heaviside function. The integral in (35) can be done analitically (see the Appendix). The result for $\rho_{x}$ is:

\begin{equation}
\rho_{x}(t, \vec x) = e^{-\lambda t} \Bigg{[} \delta(x-vt) \delta(y) + 
{\lambda \over{2 \pi v (vt-x)}} \exp \Big{(} {{\lambda \over v} \sqrt {(vt)^{2}-r^2}} \Big{)}  H(vt-r)\Bigg{]}.
\end{equation}

$\rho_{x}(t, \vec x)$ is, in 2 dimensions, the probability density for finding the particle in $\vec x$ at time $t$ provided that at time $t=0$ it left the origin moving to the right. As long as the particle doesn't collide its $\rho_{x}(t, \vec x)$ will be $\delta (vt-x) \ \delta (y)$, where $v$ is the speed of the particle. Since the particle does suffer, on the average, $\lambda$ collisions per unit time, the density of probability contributed by the possibility of no scattering is $e^{-\lambda t} \ \delta (vt-x) \ \delta (y)$. Once the particle is scattered at a time $t$' and to a direction $\varphi'$ - which will happen with a probability density $\lambda f(\varphi') e^{-\lambda t'}$ - its contribution to $\rho_{x}(t, \vec x)$ will be $\rho_{x}(t-t', R_{-\varphi'}(\vec x-vt'\vec i))$, where $R_{-\varphi'}$ is the rotation that takes the $\varphi'$ direction back to the positive $x$-axis direction and $\vec i$ is the unit vector that points to the right. All this paragraph is the following integral equation:

\begin{equation}
\rho_{x}(t, \vec x) = e^{-\lambda t} \ \delta (vt-x) \ \delta (y) + 
\lambda\ \int_{0}^{t} dt'\ e^{-\lambda t'} \int_{0}^{2 \pi} d\varphi' \ f(\varphi') \ 
\rho_{x}(t-t', R_{-\varphi'}(\vec x-vt'\vec i))
\end{equation}

The Fourier transform that we use for continuous variable is:

\begin{equation}
\int_{- \infty}^{\infty} dx\ {\rm{function}}(x)\ e^{-i 2 \pi \nu x},
\end{equation}

\noindent
and the same transformation except for the sign of the exponent for its inverse. The Fourier transform of the integral equation is:

\begin{equation}
\tilde{\rho_{x}}(t, \vec \nu) = e^{-\lambda t} e^{-i 2 \pi (\vec \nu)_{x} v t} + \lambda\ \int_{0}^{t} dt' \ \int_{0}^{2 \pi} d\varphi' \ f(\varphi') \ e^{-\lambda t'} e^{-i 2 \pi (\vec \nu)_{x} v t'} \tilde{\rho_{x}}(t-t', R_{-\varphi'}\vec \nu),
\end{equation}

\noindent
where $(\vec \nu)_{x}$ stands for the $x$ component of $\vec \nu$. We now take the Laplace transform (denoted by a hat) and obtain:

\begin{equation}
(\omega + \lambda + i 2 \pi (\vec \nu)_{x} v)\ \hat{\tilde{\rho_{x}}}(\omega, \vec \nu) = 
1 + \lambda\ \int_{0}^{2 \pi} d\varphi'\ f(\varphi')\ \hat{\tilde{\rho_{x}}}(\omega, R_{-\varphi'} \vec \nu),
\end{equation}

\noindent
where $\omega$ is the Laplace conjugate variable to time. 

Consider the case of isotropic scattering ($f(\varphi')={1 \over 2 \pi}$). Then the integral in the preceding equation is the Fourier-Laplace transform of the solution to the problem of isotropic scattering with isotropic initial conditions. For this case the following integral equation can be written for its probability density $\rho$ [8]:

\begin{equation}
\rho(t, \vec x) = \eta (t, \vec x) + \lambda \int_{0}^{t} dt'\ \int d\vec x'\ \eta (t', \vec x')\ \rho(t-t', \vec x - \vec x')
\end{equation}

\noindent
where

\begin{equation}
\eta (t, \vec x) = {e^{-\lambda t} \over 2 \pi |\vec x|}\ \delta(|\vec x|-vt).
\end{equation}

Using the Fourier and Laplace convolution theorems one can solve for the  Fourier-Laplace transform of $\rho$ [8]:

\begin{equation}
\hat{\tilde{\rho}} = {\hat{\tilde{\eta}} \over {1-\lambda \hat{\tilde{\eta}}}} \end{equation}

\noindent
in terms of the Fourier-Laplace transform of $\eta$:

\begin{equation}
\hat{\tilde{\eta}} (\omega, \vec \nu) = {1 \over \sqrt{(\lambda + \omega)^2 + (2 \pi v \nu)^2}}
\end{equation}

\noindent
Substitution into (24) yields:

$$
\hat{\tilde{\rho_{x}}}(\omega, \vec \nu) = {1 \over 
(\omega + \lambda + i 2 \pi (\vec \nu)_{x} v)\ (1-\lambda \hat{\tilde{\eta}}(\omega, \vec \nu))} = 
$$
\medskip

\begin{equation}
{\sqrt{(\lambda + \omega)^2 + (2 \pi v \nu)^2} \over 
(\omega + \lambda + i 2 \pi (\vec \nu)_{x} v)\ 
\Big{(}\sqrt{(\lambda + \omega)^2 + (2 \pi v \nu)^2} - \lambda\Big{)}},
\end{equation}

\noindent
which is the Fourier-Laplace transform of (37).

In 3 dimensions the analogous procedure starts from the equation:

\begin{equation}
\rho_{z}(t, \vec x) = e^{-\lambda t} \ \delta (x) \ \delta (y) \ \delta (vt-z) + \lambda\ \int_{0}^{t} dt'\ e^{-\lambda t'} \int d\Omega' \ f(\theta') \  \rho_{z}(t-t', R_{-\Omega'}(\vec x-vt'\vec k)).
\end{equation}

\noindent
We are assuming here that the scattering is independent of the azimuthal angle $\varphi$. $R_{-\Omega'}$ is the rotation which maps the $\Omega'$ direction into the z-direction and has its axis in the $xy$-plane.  The scattering cross section is normalized as follows:

\begin{equation}
\int d\Omega \ f(\theta) = 2 \pi \int_{0}^{\pi} d\theta\ \sin \theta\ f(\theta) = 1.
\end{equation}

\noindent
The Fourier-Laplace transform of the integral equation is now:

\begin{equation}
(\omega + \lambda + i 2 \pi (\vec \nu)_{z} v)\ \hat{\tilde{\rho_{z}}}(\omega, \vec \nu) = 
1 + \lambda\ \int d\Omega' f(\theta')\ \hat{\tilde{\rho_{z}}}(\omega, R_{-\Omega'} \vec \nu),
\end{equation}
\medskip

Consider the case of isotropic scattering ($f(\theta)={1 \over 4 \pi }$). Then the integral in the preceding equation is the Fourier-Laplace transform of the solution to the problem of isotropic scattering with isotropic initial conditions. For the this case the integral equation (42) holds and its solution is again given by (44). The difference is that $\eta$ is now:

\begin{equation}
\eta (t, \vec r) = {e^{-\lambda t} \over 4 \pi r^2}\ \delta(r-vt), 
\end{equation}

\noindent
and $\hat{\tilde{\eta}} (\omega, \vec \nu)$ is:

\begin{equation}
\hat{\tilde{\eta}} (\omega, \vec \nu) = {1 \over {2 \pi v \nu}} {\rm{arc\ tg}} {2 \pi v \nu \over {\lambda + \omega}}.
\end{equation}

\noindent
Finally, the Fourier-Laplace transform of $\hat{\tilde{\rho_{z}}}(\omega, \vec \nu)$ in 3 dimensions for the case of isotropic scattering is:

\begin{equation}
\hat{\tilde{\rho_{z}}}(\omega, \vec \nu) = {1 \over 
(\omega + \lambda + i 2 \pi (\vec \nu)_{x} v)}\ {2 \pi v \nu  \over {2 \pi v \nu - \lambda\ {\rm{arc\ tg}} {2 \pi v \nu \over {\lambda + \omega}}}}.
\end{equation}

\section{Anisotropic scattering}

The form of the integral in (41) is similar to a convolution. This suggests the following procedure to factorize the unknown function out of the integral. First, write $\vec \nu$ as $R_{\varphi} \vec \nu$:

\begin{equation}
(\omega + \lambda + i 2 \pi (R_{\varphi}\vec \nu)_{x} v)\ \hat{\tilde{\rho_{x}}}(\omega, R_{\varphi}\vec \nu) = 
1 + \lambda\ \int_{0}^{2 \pi} d\varphi'\ f(\varphi')\ \hat{\tilde{\rho_{x}}}(\omega, R_{\varphi-\varphi'} \vec \nu).
\end{equation}

Define now the following integral transform for a function $g$ of $\vec x$ ($\vec x$ is a 2-dimensional vector):

\begin{equation}
\overline{g}(\vec x, \overline{\varphi}) \equiv \int_{0}^{2 \pi} d\varphi\ g(R_{\varphi}\vec x)\ e^{-i \overline{\varphi} \varphi},\ \ \ \ \ \ \overline{\varphi}\epsilon Z.
\end{equation}

\noindent
From Fourier series we know that this transform can be inverted:

\begin{equation}
g(R_{\varphi}\vec x) =  {1 \over {2 \pi}}\ \sum_{\overline{\varphi}\epsilon Z} \overline{g}(\vec x, \overline{\varphi})\ e^{i \overline{\varphi} \varphi}.
\end{equation}

\noindent
It has the following property:

\begin{equation}
\overline{g}(R_{\alpha} \vec x, \overline{\varphi}) = e^{i \alpha \overline{\varphi}}\ \overline{g}(\vec x, \overline{\varphi})
\end{equation}

\noindent
Thus the transform that we have defined is, up to a phase factor, the Fourier transform of the restriction of the function to a circumference about the origin.

Application of $\int_{0}^{2 \pi} d\varphi\ e^{-i \overline{\varphi} \varphi}$ to both sides of the integral equation (53) yields:

\begin{equation}
2 \pi \delta_{0, \overline{\varphi}} + \lambda \tilde{f}(\overline{\varphi}) \overline{\hat{\tilde{\rho}}}_{x}(\omega, \vec \nu, \overline{\varphi}) = 
(\omega + \lambda)\ \overline{\hat{\tilde{\rho}}}_{x}(\omega, \vec \nu, \overline{\varphi}) + 
i 2 \pi v \int_{0}^{2 \pi}d\varphi\ \hat{\tilde{\rho_{x}}}(\omega, R_{\varphi}\vec \nu)\ (R_{\varphi}\vec \nu)_{x}\ e^{-i \overline{\varphi} \varphi}.
\end{equation}

\noindent
Let $\varphi_{\vec \nu}$ denote the angle between $\vec \nu$ and the $x$ axis. Then the integral in the above equation can be developped as follows:

\begin{equation}
\int_{0}^{2 \pi}d\varphi\ \hat{\tilde{\rho_{x}}}(\omega, R_{\varphi}\vec \nu)\ (R_{\varphi}\vec \nu)_{x}\ e^{-i \overline{\varphi} \varphi} = 
$$

$$
\int_{0}^{2 \pi}d\varphi\ \hat{\tilde{\rho_{x}}}(\omega, R_{\varphi}\vec \nu)\ |\vec \nu|\ 
{
e^{i (\varphi +  \varphi_{\vec \nu})} + e^{-i (\varphi +  \varphi_{\vec \nu})} 
\over 2}\ e^{-i \overline{\varphi} \varphi} = 
$$

$$
{|\vec \nu| \over 2}\  \Big{[} e^{i \varphi_{\vec \nu}} 
\overline{\hat{\tilde{\rho}}}_{x}(\omega, \vec \nu, \overline{\varphi} - 1) + 
e^{-i \varphi_{\vec \nu}} 
\overline{\hat{\tilde{\rho}}}_{x}(\omega, \vec \nu, \overline{\varphi} + 1) \Big{]}.
\end{equation}

\noindent
Substitution of this result back into (57) yields a set of recursive equations for the transform $\overline{\hat{\tilde{\rho}}}_{x}$ of the probability density:

\begin{equation}
2 \pi \delta_{0, \overline{\varphi}} + 
[\lambda (\tilde{f}(\overline{\varphi}) - 1) - \omega]\ \overline{\hat{\tilde{\rho}}}_{x}(\omega, \vec \nu, \overline{\varphi}) = 
$$

$$
i \pi v |\vec \nu|\ 
\Big{[} e^{i \varphi_{\vec \nu}} 
\overline{\hat{\tilde{\rho}}}_{x}(\omega, \vec \nu, \overline{\varphi} - 1) + 
e^{-i \varphi_{\vec \nu}} 
\overline{\hat{\tilde{\rho}}}_{x}(\omega, \vec \nu, \overline{\varphi} + 1) \Big{]},\ \ \ \ \ \ \ \ \overline{\varphi}\epsilon Z.
\end{equation}

If the scattering has left-right symmetry ($f(\varphi) = f(2 \pi- \varphi)$), then, 

\begin{equation}
\tilde{f}(\overline{\varphi}) = \tilde{f}(-\overline{\varphi})
\end{equation}

\noindent
For $\overline{\hat{\tilde{\rho}}}_{x}$ this symmetry implies:

\begin{equation}
\overline{\hat{\tilde{\rho}}}_{x}(\omega, |\vec \nu| \vec i, -\overline{\varphi}) =
\overline{\hat{\tilde{\rho}}}_{x}(\omega, |\vec \nu| \vec i, \overline{\varphi}). 
\end{equation}

\noindent
From this and (56):

\begin{equation}
\overline{\hat{\tilde{\rho}}}_{x}(\omega, \vec \nu, \pm \overline{\varphi}) =
e^{\pm i \overline{\varphi} \varphi_{\vec \nu}}\ \overline{\hat{\tilde{\rho}}}_{x}(\omega, |\vec \nu| \vec i, \overline{\varphi}). 
\end{equation}

Now the set of equations (59) can be simplified to:

\begin{equation}
2 \pi - \omega \overline{\hat{\tilde{\rho}}}_{x}(\omega, |\vec \nu|, 0) = 
i 2 \pi v |\vec \nu|\ \overline{\hat{\tilde{\rho}}}_{x}(\omega, |\vec \nu| \vec i, 1)
\end{equation}

\noindent
and

$$
[\lambda (\tilde{f}(\overline{\varphi}) - 1) - \omega]\  
\overline{\hat{\tilde{\rho}}}_{x}(\omega, |\vec \nu| \vec i , \overline{\varphi}) = 
i \pi v |\vec \nu|\
\Big{[} 
\overline{\hat{\tilde{\rho}}}_{x}(\omega, |\vec \nu| \vec i, \overline{\varphi} - 1) + 
\overline{\hat{\tilde{\rho}}}_{x}(\omega, |\vec \nu| \vec i, \overline{\varphi} + 1)
\Big{]}, 
$$

\begin{equation}
\rm{for}\ \ \ \overline{\varphi} = 1, 2,...,
\end{equation}

\noindent
or (bearing (61) in mind and remembering that $\tilde{f}(0)=1$) simply to

$$
[\lambda (1-\tilde{f}(\overline{\varphi})) + \omega]\  
\overline{\hat{\tilde{\rho}}}_{x}(\omega, |\vec \nu| \vec i , \overline{\varphi}) + 
i \pi v |\vec \nu|\
\Big{[} 
\overline{\hat{\tilde{\rho}}}_{x}(\omega, |\vec \nu| \vec i, \overline{\varphi} - 1) + 
\overline{\hat{\tilde{\rho}}}_{x}(\omega, |\vec \nu| \vec i, \overline{\varphi} + 1)
\Big{]} = 2 \pi \delta_{0, \overline{\varphi}}, 
$$

\begin{equation}
\rm{for}\ \ \ \overline{\varphi} = 0, 1, 2,....
\end{equation}

\noindent
This more compact form brings out the matricial structure of the equations: 

$$
\left(
\begin{array}{ccccc}
\cdot & \cdot & \cdot & \cdot & \cdot \\
\cdot & i \pi v |\vec \nu| & [\lambda (1-\tilde{f}(2)) + \omega] & i \pi v |\vec \nu| & 0 \\
\cdot & 0 & i \pi v |\vec \nu| & [\lambda (1-\tilde{f}(1)) + \omega] & i \pi v |\vec \nu| \\
\cdot & 0 & 0 & 2 i \pi v |\vec \nu| & [\lambda (1-\tilde{f}(0)) + \omega] 
\end{array}
\right)
\left(
\begin{array}{c}
\cdot \\
\overline{\hat{\tilde{\rho}}}_{x}(\omega, |\vec \nu| \vec i , 2) \\
\overline{\hat{\tilde{\rho}}}_{x}(\omega, |\vec \nu| \vec i , 1) \\
\overline{\hat{\tilde{\rho}}}_{x}(\omega, |\vec \nu| \vec i , 0)
\end{array}
\right)
$$

\begin{equation}
=
\left(
\begin{array}{c}
\cdot \\
0 \\
0 \\
2 \pi
\end{array}
\right)
\end{equation}

From (55), (61) and (62) the Fourier-Laplace transform of the probability density is: 

\begin{equation}
\hat{\tilde{\rho_{x}}}(\omega, \vec \nu) =  {1 \over {\pi}}\ \Bigg{(} {\overline{\hat{\tilde{\rho}}}_{x}(\omega, |\vec \nu| \vec i, 0)\  \over 2}+ \sum_{\overline{\varphi} = 1,2,..}\ \overline{\hat{\tilde{\rho}}}_{x}(\omega, |\vec \nu| \vec i, \overline{\varphi})\  \cos ({\overline{\varphi}}\ \varphi_{\vec \nu})\ \Bigg{)}
\end{equation}

\noindent
Thus by approximating equation (66) by a finite dimensional truncation we can find the Fourier-Laplace transform of the probability density with arbitrary precision. There are recursive formulae that allow to invert finite tridiagonal matrices at a low computational cost [18]. 

We assume that at $t=0$ all particles leave from the same point and have the same speed. If this is not the case, summation over speeds and positions is necessary. Let $I(\varphi)={1 \over 2 \pi} \sum_{{\overline{\varphi}} \epsilon Z} {\tilde{I}} (\overline{\varphi})\ e^{i {\overline{\varphi}} \varphi}$ be the initial angular distribution of velocities. Let ${\overline{\hat{\tilde{\rho}}}}_{\varphi}(\omega, \vec \nu, {\overline{\varphi}})$ be the function corresponding to ${\overline{\hat{\tilde{\rho}}}}_{x}(\omega, \vec \nu, {\overline{\varphi}})$ when the initial direction of the particle is $\varphi$. $R_{-\varphi}$ is the adjoint operator of $R_{\varphi}$, therefore:

\begin{equation}
{\overline{\hat{\tilde{\rho}}}}_{\varphi}(\omega, \vec \nu, {\overline{\varphi}}) = {\overline{\hat{\tilde{\rho}}}}_{x}(\omega,R_{-\varphi} \vec \nu, {\overline{\varphi}}).
\end{equation}

\noindent
Then, using standard properties of Fourier sums and property (56) of the\ ${\bar{}}$\ transform,

\begin{equation}
\int_{0}^{2 \pi} d \varphi\ I(\varphi)\ {\overline{\hat{\tilde{\rho}}}}_{\varphi}(\omega, \vec \nu, {\overline{\varphi}}) = {\tilde{I}}({\overline{\varphi}})\  {\overline{\hat{\tilde{\rho}}}}_{x}(\omega, \vec \nu, {\overline{\varphi}}).
\end{equation}

In 3 dimensions we can proceed analogously. The starting equation is now (49).
In it we write $\vec \nu$ as $R_{\Omega} \vec \nu$. 

\begin{equation}
(\omega + \lambda + i 2 \pi (R_{\Omega} \vec \nu)_{z} v)\ \hat{\tilde{\rho_{z}}}(\omega, R_{\Omega} \vec \nu) = 
1 + \lambda\ \int d\Omega' f(\theta')\ \hat{\tilde{\rho_{z}}}(\omega, R_{\Omega-\Omega'} \vec \nu),
\end{equation}

\noindent
3-dimensional rotations can not be added commutatively, but $R_{\Omega-\Omega'}$ is well defined. It is the rotation that maps the $\Omega'$ direction into the $\Omega$ direction and has its axis perpendicular to the plane defined by these directions and the origin. 

Define now the following integral transform for a function $g$ of $\vec x$ ($\vec x$ is a 3-dimensional vector):

\begin{equation}
\overline{g}(\vec x, l) \equiv \int d\Omega\ g(R_{\Omega}\vec x)\ Y_{l}^{0}(\Omega),\ \ \ \ \ \ \ l\epsilon N.
\end{equation}

\noindent
If $g$ does not depend on the azimuthal angle $\varphi$ the following 2 properties hold:

\noindent
a)

\begin{equation}
\overline{g}(R_{\Omega} \vec x, l) =  \overline{g}(\vec x, l) \sqrt{4 \pi \over {2 l + 1}}\ Y_{l}^{0}(\Omega) 
\end{equation}

\noindent
and b) the above transform is one-to-one and its inverse is:

\begin{equation}
g(\vec x) =  \sum_{l \epsilon N} \overline{g}(\vec x, l)\ Y_{l}^{0*}(0,0) =  \sum_{l \epsilon N} \sqrt{2l+1 \over 4 \pi}\ \overline{g}(\vec x, l),
\end{equation}

\noindent
where $(0,0)$ means $(\theta, \varphi)=(0,0)$. Similarly to (10) we define:

\begin{equation}
f_{l} = \sqrt{{4 \pi \over 2l+1}}\ \int d\Omega\ f(\Omega) Y_{l}^{0} (\Omega).
\end{equation}

\noindent
Application of $\int d\Omega\ Y_{l}^{0}(\Omega)$ to both sides of the integral equation (70) and use of the convolution theorem (13) yields:

\begin{equation}
(\omega + \lambda)\ \overline{\hat{\tilde{\rho}}}_{z}(\omega, \vec \nu, l) + 
i 2 \pi v \int d\Omega\ \hat{\tilde{\rho_{z}}}(\omega, R_{\Omega}\vec \nu)\ (R_{\Omega}\vec \nu)_{z}\ Y_{l}^{0}(\Omega)
=\sqrt{4 \pi} \delta_{0, l} + \lambda f_{l} \overline{\hat{\tilde{\rho}}}_{z}(\omega, \vec \nu, l).
\end{equation}

\noindent
If $\Omega \equiv (\theta, \varphi)$ and the polar coordinates of $\vec \nu$ are $(\theta_{\vec \nu}, \varphi_{\vec \nu})$, then geometry yields:

\begin{equation}
(R_{\Omega}\vec \nu)_{z} = |\vec \nu|\ (\cos \theta \cos \theta_{\vec \nu} - \sin \theta \sin \theta_{\vec \nu} \cos (\varphi-\varphi_{\vec \nu})).
\end{equation}

\noindent
Due to the assumed cilindrical symmetry of the scattering cross section we can take $\varphi_{\vec \nu}=0$. Then, writing the trigonometric functions as linear combinations of spherical harmonics, we obtain:

\begin{equation}
(R_{\varphi}\vec \nu)_{z} = \sqrt{4 \pi \over 3}\ |\vec \nu|\ \Big{[} Y_{1}^{0}(\Omega) \cos \theta_{\vec \nu} - \big{(} Y_{1}^{-1}(\Omega) - Y_{1}^{1}(\Omega) \big{)} {\sin \theta_{\vec \nu} \over \sqrt{2}} \Big{]}.
\end{equation}

When we substitute this expression back into (75) we are going to have products of spherical harmonics. By means of Clebsch-Gordan coefficients [19] the products $Y_{1}^{\pm 1}(\Omega)\ Y_{l}^{0}(\Omega)$ can be written as linear combinations of $Y_{l \pm 1}^{\pm 1}(\Omega)$. But it easy to check that for a cilindrically symmetric function $g(\vec x),\ \int d\Omega\ g(R_{\Omega}\vec x)\ Y_{l}^{m}(\Omega) = 0$ if $m \neq 0$. Therefore only the product $\sqrt{4 \pi \over 3}\ Y_{l}^{0}(\Omega)\ Y_{1}^{0}(\Omega)$ remains. Using the recurrence relations between Legendre polynomials [10] this product can be expanded as follows:

\begin{equation} 
\sqrt{4 \pi \over 3}\ Y_{l}^{0}(\Omega)\ Y_{1}^{0}(\Omega) = 
{l+1 \over \sqrt{(2l+1)(2l+3)}}\ Y_{l+1}^{0}(\Omega)+ {l \over \sqrt{(2l+1)(2l-1)}}\ Y_{l-1}^{0}(\Omega).
\end{equation}

\noindent
Substituting this result back into (75) we obtain the following set of recursive equations for $\overline{\hat{\tilde{\rho}}}_{z}$:

$$
[\lambda (1-f_{l}) + \omega]\  
\overline{\hat{\tilde{\rho}}}_{z}(\omega, \vec \nu , l) + 
{i 2 \pi v |\vec \nu| (l+1) \over \sqrt{(2l+1)(2l+3)}} \cos \theta_{\vec \nu}\ \overline{\hat{\tilde{\rho}}}_{z}(\omega, \vec \nu, l+1) + 
{i 2 \pi v |\vec \nu| l \over \sqrt{(2l+1)(2l-1)}} \cos \theta_{\vec \nu}\ \overline{\hat{\tilde{\rho}}}_{z}(\omega, \vec \nu, l-1)
$$

$$
= \sqrt{4 \pi} \delta_{0, l}, 
$$

\begin{equation}
{\rm{for}}\ \ \ l = 0, 1, 2,....
\end{equation}

\noindent
In matricial form: 

$$
\left(
\begin{array}{ccccc}
\cdot & \cdot & \cdot & \cdot & \cdot \\
\cdot & {i 2 \pi v |\vec \nu| 3 \over \sqrt{35}} \cos \theta_{\vec \nu} & [\lambda (1-f_{2}) + \omega] & {i 2 \pi v |\vec \nu| 2 \over \sqrt{15}} \cos \theta_{\vec \nu} & 0 \\
\cdot & 0 & {i 2 \pi v |\vec \nu| 2 \over \sqrt{15}} \cos \theta_{\vec \nu} & [\lambda (1-f_{1}) + \omega] & {i 2 \pi v |\vec \nu| \over \sqrt{3}} \cos \theta_{\vec \nu} \\
\cdot & 0 & 0 & {i 2 \pi v |\vec \nu| \over \sqrt{3}} \cos \theta_{\vec \nu} & [\lambda (1-f_{0}) + \omega] 
\end{array}
\right)
\left(
\begin{array}{c}
\cdot \\
\overline{\hat{\tilde{\rho}}}_{z}(\omega, \vec \nu, 2) \\
\overline{\hat{\tilde{\rho}}}_{z}(\omega, \vec \nu, 1) \\
\overline{\hat{\tilde{\rho}}}_{z}(\omega, \vec \nu, 0)
\end{array}
\right)
$$

\begin{equation}
=
\left(
\begin{array}{c}
\cdot \\
0 \\
0 \\
\sqrt{4 \pi},
\end{array}
\right)
\end{equation}

\noindent
where, for symmetry's sake, we have not simplified $[\lambda (1-f_{0}) + \omega]$ to $\omega$.

Truncation  of the equations to a finite $l$ yields a set of equations that can be solved. As in the 2-dimensional case the matrix is tridiagonal and it can be inverted using recursive formulae [18]. By means of the inversion formula (73) we can thus find the Fourier-Laplace transform of the probability density with arbitrary precision.

As in the 2-dimensional case we assume that at $t=0$ all particles leave from the same point and have the same speed. Let $I(\Omega)= \sum_{l,m} I_{l,m} Y_{l}^{m}(\Omega)$ be the initial angular distribution of velocities. Let ${\overline{\hat{\tilde{\rho}}}}_{\Omega}(\omega, \vec \nu, l)$ be the function corresponding to ${\overline{\hat{\tilde{\rho}}}}_{z}(\omega, \vec \nu, l)$ when the initial direction of the particle is $\Omega$. $R_{-\Omega}$ is the adjoint operator of $R_{\Omega}$, therefore:

\begin{equation}
{\overline{\hat{\tilde{\rho}}}}_{\Omega}(\omega, \vec \nu, l) = {\overline{\hat{\tilde{\rho}}}}_{z}(\omega,R_{-\Omega} \vec \nu, l).
\end{equation}

\noindent
Then, using standard properties of spherical harmonics and property (72) of the \ ${\bar{}}$\ transform,

\begin{equation}
\int d \Omega\ I(\Omega)\ {\overline{\hat{\tilde{\rho}}}}_{\Omega}(\omega, \vec \nu, l) = \sqrt{4 \pi \over {2 l + 1}}\ {\overline{\hat{\tilde{\rho}}}}_{z}(\omega, \vec \nu, l)\ I_{l,0}.
\end{equation}

This result does not depend on the value of the $I_{l,m}$'s for $m \neq 0$. As noted earlier $\int d \Omega\ I(\Omega) {\overline{\hat{\tilde{\rho}}}}_{\Omega}(\omega, \vec \nu, l)$ can be inverted by means of (73) to yield the Fourier-Laplace transform of $\rho(t, \vec x)$ only when $I$ is cilidrically symmetric. Thus to apply formula (82) one has to decompose a general initial condition as a sum of cilindrically symmetric (with respect to different axis) components and then apply (82) to each component separately.

\section*{Acknowledgments}

Conversations with Andr\'es Hombr\'\i a\ Mat\'e are happily remembered.

\bigskip

\centerline{}

\bigskip

\noindent
{\Large{\bf{Appendix}}}
\bigskip

Substitution of (36) into (35) yields 3 terms, which are:

\begin{equation}
e^{-\lambda t} \delta(x-vt) \delta(y),
\end{equation}

\begin{equation}
\lambda e^{-\lambda t} \int_{0}^{t} dt'\ {\delta(|\vec x - v t'\vec i|-v(t-t')) \over 2 \pi |\vec x - v t'\vec i|} 
\end{equation}

\noindent
and

\begin{equation}
\lambda^2 e^{-\lambda t} \int_{0}^{t} dt'\ { \exp \Big{(} {{\lambda \over v} \sqrt {(v(t-t'))^{2}-|\vec x - v t'\vec i|^2}} \Big{)}  \over{2 \pi v  \sqrt {(v(t-t'))^{2}-|\vec x - v t'\vec i|^2}}} H(v(t-t')-|\vec x - v t'\vec i|).
\end{equation}

Since the integration variable is not the argument of the Dirac delta function in the second term, the following formula will be needed:

\begin{equation}
\int dx\ \delta(u(x)) = \sum_{u(x)=0} {1 \over |u'(x)|}.
\end{equation}

\noindent
It is convenient to write the integral in the second term (84) as follows:

\begin{equation}
{1 \over 2 \pi v} \int_{0}^{vt} dy\ {\delta(|\vec x - y \vec i|-vt+y)) \over 2 \pi |\vec x - y \vec i|}. 
\end{equation}

\noindent
The derivative that appears in the r.h.s. of (86) is then:

\begin{equation}
{\partial \over {\partial y}}(|\vec x - y \vec i|-vt+y)= 1+ {y-x \over |\vec x - y \vec i|}. 
\end{equation}

\noindent
Using of the last 3 formulae the second term becomes 

\begin{equation}
{\lambda e^{-\lambda t} \over {2 \pi v}} {1 \over {y-x+|\vec x - y \vec i|}}
\end{equation}

\noindent
The argument of the Dirac delta function in the second term vanishes for

\begin{equation}
y= {v^2 t^2 - r^2 \over{2 (vt-x)}}. 
\end{equation}

\noindent
Substituting this in (88) yields

\begin{equation}
{\lambda e^{-\lambda t} \over {2 \pi v}} {1 \over {vt-x}}.
\end{equation}

Defining 

\begin{equation}
A \equiv v^2 t^2 - r^2 
\end{equation}

\noindent
and

\begin{equation}
B \equiv 2 v (x-vt),
\end{equation}

\noindent
the third term (85) can be written as

$$
{\lambda^2 e^{-\lambda t} \over 2 \pi v} \int_{0}^{{v^2 t^2 - r^2 \over{2 (vt-x)}}} dt'\ \sum_{j=0}^{\infty} {\lambda^j (A+B t')^{j-1 \over {2}}\over{j!\ v^j}}=
$$

$$
{\lambda e^{-\lambda t} \over 2 \pi v (x-vt)} \Bigg{[} \exp {\lambda \sqrt{A+B t'} \over {v}} -1 \Bigg{]}_{0}^{v^2 t^2 - r^2 \over{2 (vt-x)}}=
$$

\begin{equation}
{\lambda e^{-\lambda t} \over 2 \pi v (vt-x)} \Bigg{[} \exp \Bigg{(} {\lambda \sqrt{v^2 t^2 - r^2 } \over {v}} \Bigg{)} -1 \Bigg{]}.
\end{equation}

Collecting results we obtain (37).

\vfill

\eject

\section*{References}

\begin{verse}


[1]A. V. Starkov, M. Noormohammadian, U. G. Oppel, "A stochastic model and a variance-reduction Monte-Carlo method for the calculation of light transport", {\it Applied Physics B, Lasers and Optics}, B 60, (1995) 335-340.

[2] Milton Kerker, "The scattering of light" (Academis Press, 1969).

[3] R. F. Bonner and R. Nossal, Appl. Opt. 20 (1981) 2097.

[4] A. P. Sheperd and P. A. Oberg, Laser-Doppler Blood Flowmetry (Kluwer, New York, 1989)

[5] B. Chance {\it{et alii}}, Pro. Natl. Acad. Sci. USA 85 (1988) 4971.

[6] J. M. Fern\'andez-Varea {\it{et alii}}, Nucl. Instr. and Meth. in Phys. Res. B73 (1993), 447-473.

[7] R. Garc\'{\i}a-Pelayo, Physica A, 216, (1995) 299-315.

[8] G. I. Bells and S. Glasstone, "Nuclear Reactor Theory" (Van Nostrand Reinhold, 1970)

[9] J. H. Ferziger {\it{et al.}}, "The theory of neutron slowing down in nuclear reactors" (Pergamon Press, New York, 1966).

[10] Harry Hochstadt, "The Functions of Mathematical Physics", {\it Dover} (1986). 

[11] Chun Wa Wong, "Introduction to Mathematical Physics", {\it Oxford} (1991).

[12] S. Goudsmit and J. L. Saunderson, Phys. Rev. 57 (1940) 24.

[13] S. Goudsmit and J. L. Saunderson, Phys. Rev. 58 (1940) 36.

[14] H. W. Lewis, Phys. Rev. 78 (1950) 526.

[15] C. S. Johnson Jr., Don A. Gabriel, "Laser light scattering", {\it Dover}, (1995) p. 53.

[16] M. Bogu\~{n}\'a {\it{et alii}}, Physica A 230 (1996) 149-155.

[17] J. Masoliver, J. M. Porr\'a and G. H. Weiss, Physica A 193 (1993) 469-482.

[18] Y. Huang and W. F. McColl, J. Phys. A (1997) 7919-7933.

[19] C. Cohen-Tannoudji, B. Diu and F. Lalo\"e, "Quantum Mechanics", {\it Wiley-Interscience} (1977).

\end{verse}

\end{document}